\pdfoutput=1 
\documentclass[prd,singlecolumn,preprintnumbers,superscriptaddress,showpacs,preprint]{revtex4}
\usepackage{graphicx}
\usepackage{bm}
\usepackage{amsmath}
\usepackage{amssymb}

\newcommand\be{\begin{eqnarray}}
\newcommand\ee{\end{eqnarray}}

\begin{document}


\author{Dmitri E. Kharzeev}
\affiliation{Center for Nuclear Theory, Department of Physics and Astronomy, Stony Brook University, New York 11794-3800, USA}
\affiliation{Department of Physics and RIKEN-BNL Research Center, Brookhaven National Laboratory, Upton, New York 11973-5000, USA}

\title{The mass radius of the proton}

\begin{abstract}
The mass radius is a fundamental property of the proton that so far has not been determined from experiment. Here we show that the mass radius of the proton can be rigorously defined through the formfactor of the trace of the energy-momentum tensor (EMT) of QCD in the weak gravitational  field approximation, as appropriate for this problem.   We then demonstrate that the scale anomaly of QCD enables the extraction of the formfactor of the trace of the EMT from the data on threshold photoproduction of $J/\psi$ and $\Upsilon$ quarkonia, and use the recent GlueX Collaboration data to extract the r.m.s. mass radius of the proton ${\rm R_m = 0.55 \pm 0.03 \ fm}$. The extracted mass radius is significantly smaller than the r.m.s. charge radius of the proton ${\rm R_C = 0.8409 \pm 0.0004 \ fm}$. We attribute this difference to the interplay of asymptotic freedom and spontaneous breaking of chiral symmetry in QCD, and outline future measurements needed to determine the mass radius more precisely.
 \end{abstract}

\pacs{12.38.Aw;12.40.Yx;13.60.Le}
\maketitle

\section{Introduction}

The mass distribution is a fundamental property of a physical object. Yet, while a lot of information is available about the charge distribution inside the proton, nothing is known at present about its mass radius. In astrophysics and cosmology, the study of the mass distribution in galaxies has led to establishing the presence of Dark Matter that is believed to constitute about $85 \%$ of the total mass of matter in the Universe. Drawing an analogy to the physics of the proton, the electron scattering experiments reveal the spatial distribution of quarks (matter visible to photons), but do not directly  constrain the spatial distribution of gluons -- ``dark matter of QCD" that is not visible to photons.  One may thus fully expect that an experimental determination of the mass distribution would constitute a big advance in the understanding of the proton structure. 
\vskip0.3cm

Because of the extreme weakness of the gravitational field created by a single proton, its direct measurement at short distances is clearly impossible. Likewise, a study of graviton--proton scattering is off limits for present experiments. Does this mean that the mass radius of the proton cannot be measured? We believe that the possibility to measure the mass distribution inside the proton is provided by the scale anomaly, reflecting the asymptotic freedom of QCD \cite{Gross:1973id,Politzer:1973fx}. In the chiral limit of massless quarks, the scale anomaly \cite{Ellis:1970yd,Chanowitz:1972vd} expresses the trace of the energy-momentum tensor (EMT) of QCD in terms of the scalar gluon operator \cite{Collins:1976yq,Nielsen:1977sy}. It has been proposed \cite{Kharzeev:1995ij,Kharzeev:1998bz} that the matrix elements of this operator (which is largely responsible for the mass of the proton) can be extracted from the photoproduction of heavy quarkonia near the threshold.
Below, we will show how the formfactor of the scalar gluon operator can be determined from the recent data on photoproduction of $J/\psi$ near the threshold recently reported by the GlueX Collaboration \cite{ali2019first}. We will then use this formfactor to extract the mass radius of the proton from the GlueX data.
\vskip0.3cm

\section{The mass distribution and gravitational formfactors}\label{sec:grav}

As a first step, let us review how the Newton's law of gravitation emerges from the Einstein theory \cite{Einstein:1916vd} in the weak gravitational field, non-relativistic approximation. The Einstein equation reads
\be\label{einstein}
R_{\mu\nu} - \frac{1}{2} g_{\mu\nu} R = 8 \pi G\ T_{\mu\nu} ,
\ee 
where $g_{\mu\mu}$ is the metric tensor, $R_{\mu\nu}$ is the Ricci curvature tensor, $R$ is the scalar curvature (Ricci scalar), $G$ is the Newton's constant, and $T_{\mu\nu}$ is the EMT. We have omitted the cosmological constant term that is not relevant for our present discussion, and put the speed of light $c=1$.

Taking the trace w.r.t. the metric on both sides of (\ref{einstein}), we get
\be\label{einstein-tr}
- R = 8 \pi G\ T ,
\ee 
where $T \equiv T_\mu^\mu$ is the trace of the EMT. This equation relates the trace of the EMT to the scalar curvature of space-time -- in fact, this relation first appeared in the extension of Nordstr\"om's scalar gravity theory \cite{Nordstrom:1912} proposed by Einstein and Fokker \cite{Einstein:1914}; see \cite{norton1984einstein} for a historical overview. In classical Maxwell electrodynamics, $T_{\rm EM} = 0$ without massive charges, so the electromagnetic field does not curve space-time, and does not gravitate. Moreover, light does not bend in the presence of massive bodies if they induce only a scalar curvature -- and thus the observation of light bending has ruled out the scalar gravity in favor of the tensor one (\ref{einstein}) proposed by Einstein \cite{Einstein:1916vd} in 1915. 

Nevertheless, in weak gravitational fields the trace of EMT and the temporal component of the EMT $T_0^0$ coincide. Therefore the distribution of mass can be obtained from the formfactor of the trace of the EMT -- we will call it for brevity the ``scalar gravitational formfactor", because this would be the only formfactor in  Nordstr\"om's scalar gravity \cite{Nordstrom:1912}. To show the equivalence of the distributions of $T$ and $T_0^0$ in a weak gravitational field, let us review how the Newtonian limit emerges from the Einstein theory, see e.g. \cite{Landau}.

In the non-relativistic limit 
\be\label{metric-nr}
g_{00} = 1 + 2 \varphi , 
\ee
where $\varphi$ is the gravitational field potential, 
and the EMT is given by
\be\label{emt-nr}
T_\mu^\nu = \mu\ u_{\mu} u^{\nu} ,
\ee
where $\mu$ is the mass density, and the 4-velocity of the non-relativistic body can be chosen as $u_0 = u^0 =1$, with all spatial components equal to zero, $u_i = 0$. Therefore, in this limit
\be\label{nr-lim}
T_0^0 = \mu; \hskip1cm T \equiv T_\mu^\mu = T_0^0 = \mu ,
\ee
so the distribution of mass and the distribution of the trace of the EMT indeed coincide. 

The equations (\ref{einstein}) and (\ref{einstein-tr}) lead to
\be
R_{\mu\nu} = 8 \pi G (T_{\mu\nu} - \frac{1}{2} g_{\mu\nu} T) ;
\ee
the temporal component $\mu=\nu=0$ of this equation is
\be\label{nr-eins}
R_0^0 = 4 \pi G \mu ,
\ee
and all other components vanish. For the metric (\ref{metric-nr}), we get
\be
R_0^0 = \frac{\partial^2 \varphi}{\partial x^{\mu 2}} \equiv \Delta \varphi ,
\ee
and thus (\ref{nr-eins}) yields the equation describing the gravitational field in Newtonian mechanics:
\be
\Delta \varphi = 4 \pi G \mu .
\ee
Its solution gives the gravitational field potential created by a distribution of mass $\mu(R)$:
\be
\varphi = - G\ \int \frac{\mu(R)\ dV}{R} .
\ee 
For a particle of mass $M = \int \mu \ dV$, the total potential is $\varphi = - GM/R$, and the force acting on a probe mass $m$ is $F_g = - m\ \partial \varphi/ \partial R$ which yields the Newton's law of gravity:
\be
F_g = - G\ \frac{m M}{R^2} .
\ee
\vskip0.3cm

The purpose of reviewing this textbook derivation was to show that in the weak gravitational field, non-relativistic, limit of gravity the distribution of mass and the distribution of the trace of the EMT are identical, see (\ref{nr-lim}). Therefore, to measure the mass distribution of a microscopic object with a weak gravitational field, instead of utilizing graviton scattering, we can measure its scalar gravitational formfactor (i.e. the formfactor of the trace of EMT). 
\vskip0.3cm

We can arrive at the same conclusion by comparing the formfactor of the EMT with the scalar gravitational formfactor. 
Let us consider first the formfactor of the EMT for a proton (spin $1/2$ particle of mass $M$) \cite{Pais:1949vdk,Pagels:1966zza}:
 \begin{eqnarray*}\label{gen}
\langle {\bf p}_1 | T_{\mu\nu} | {\bf p}_2 \rangle &=& \left( \frac{M^2}{p_{01}\ p_{02}} \right)^{1/2} \frac{1}{4 M}\ {\bar u}(p_1, s_1) \Big[  G_1(q^2) (p_\mu \gamma_\nu + p_\nu \gamma_\mu) + G_2(q^2) \frac{p_\mu p_\nu}{M} + \\
&& {} + G_3(q^2) \frac{(q^2 g_{\mu\nu} - q_\mu q_\nu)}{M} \Big] u (p_2, s_2) ,
\end{eqnarray*}
where $p_\mu = (p_1 + p_2)_\mu$, $q_\mu = (p_1 - p_2)_\mu$, $p_1^2 = p_2^2 = M^2$, and the four-component spinor  $u (p, s)$ satisfies the free Dirac equation 
$({\hat p} - M) u (p, s) = 0$ and is normalized according to $\sum_s {\bar u} (p, s) u(p, s) = ({\hat p} + M)/2M$. The formfactors $G_i(q^2)$ completely describe the mechanical structure of the spin $1/2$ particle. 

The energy-momentum conservation
\be
\partial^\mu T_{\mu\nu} = 0
\ee
implies
\be\label{Ward}
q^\mu \langle {\bf p}_1 | T_{\mu\nu} | {\bf p}_2 \rangle = 0;
\ee
it is easy to check that (\ref{gen}) satisfies the condition (\ref{Ward}) for on-shell nucleons that obey the free Dirac equation; the symmetry of (\ref{gen}) w.r.t. the interchange of $p_1$ and $p_2$ is necessary for (\ref{Ward}) to hold.

\vskip0.3cm

In the limit of vanishing momentum transfer $q_\mu \to 0$, the forward matrix element of the energy-momentum tensor takes the form 
\be\label{gen_zero}
\langle {\bf p} | T_{\mu\nu} | {\bf p} \rangle = \left( \frac{M^2}{p_0^2} \right)^{1/2} \ {\bar u}(p, s) u (p, s)\  \frac{p_\mu p_\nu}{M^2} \left[  G_1(0)  + G_2(0) \right]  ,
\ee
characterized by the $q^2=0$ values of the formfactors $G_1(q^2)$ and $G_2(q^2)$. The Hamiltonian $H$ is given by the temporal component of the energy-momentum tensor:
\be
H = \int d^3x\ T_{00} (x);
\ee
therefore, in the rest frame of the particle, the forward matrix element of $T_{00}$ should yield the mass of the particle:
\be\label{mass}
\langle {\bf p}=0 | T_{00} | {\bf p}=0 \rangle = M.
\ee
Imposing this constraint on (\ref{gen_zero}), we get the condition
\be\label{constr}
G_1(0)  + G_2(0) = M .
\ee 
The derivation of this condition is completely analogous to the derivation of the condition on the electromagnetic formfactor $F(q^2=0) = e$, where $e$ is the electric charge $e$ of the particle.
\vskip0.3cm

Let us now consider the matrix element of the trace of the EMT $T \equiv T_\mu^\mu$; from (\ref{gen}) we find
\be\label{trace-form0}
\langle {\bf p}_1 | T | {\bf p}_2 \rangle = \left( \frac{M^2}{p_{01}\ p_{02}} \right)^{1/2} \ {\bar u}(p_1, s_1)u (p_2, s_2)\ G(q^2)   ,
\ee 
with a new formfactor 
\be\label{trace-form}
G(q^2) = G_1(q^2) + G_2 (q^2) \left(1 - \frac{q^2}{4M^2}\right) + G_3 (q^2) \frac{3 q^2}{4 M^2} .
\ee
In the rest frame of the particle,
\be\label{norm-m}
\langle {\bf p}=0 | T | {\bf p}=0 \rangle = \langle {\bf p}=0 | T_{00} | {\bf p}=0 \rangle = M,
\ee
therefore 
\be\label{con-m}
G(0) = M
\ee
that is obviously consistent with (\ref{trace-form}) and (\ref{constr}).
\vskip0.3cm

\section{The mass radius}

The charge radius of the proton is usually defined \cite{Feynman:1973xc} through the derivative of its electromagnetic formfactor w.r.t. the momentum transfer $t = q^2 \equiv -Q^2$ evaluated at $t=0$:
\be
\langle R_{\rm C}^2 \rangle = - 6\ \frac{d G_{\rm EM}}{d Q^2}\Big|_{Q^2=0} .
\ee
To enable a direct comparison to the charge radius, we propose to define the mass radius analogously through the formfactor of mass density $T_{00}$ given by (\ref{gen}).
Let us compare the derivatives of $T_{00}$ and of the scalar gravitational formfactor (\ref{trace-form0}) w.r.t. $t$ at $t=0$. Because the formfactor 
of $T_{00}$ depends on the reference frame (strictly speaking, we can interpret it in terms of mass distribution only in the rest frame of the proton), we have to specify it. 

It is natural to choose the Breit frame in which ${\bf p}_2 = \frac{1}{2} {\bf q}$, ${\bf p}_1 = - \frac{1}{2} {\bf q}$. Evaluating the derivatives of the formfactors of $T_{00}$ and of the trace of the EMT w.r.t. $t$ at $t=0$, we find that they differ by terms $G_i(0)/(4M^2)$ that have to be compared to $dG_i/dt |_{t=0} \equiv G_i(0)/m_i^2$ that depend on the compositeness scales $m_i^2$ of the corresponding formfactors.

 This difference results from the frame dependence of the formfactor of $T_{00}$. Indeed,  the relativistic $\gamma$ factor for a nucleon moving with momentum ${\bf p} = \frac{1}{2} {\bf q}$ is $\gamma = E/M = \sqrt{M^2 + (q^2/4)}/M = \sqrt{1+q^2/(4M^2)}$, so for $q \equiv |{\bf q}| \simeq m_i$, the nucleon is Lorentz-contracted by $1/\gamma \simeq (1+m_i^2/(4M^2))^{-1/2}$. In 
the non-relativistic limit, when $2M \gg m_i$, this is a negligible effect -- the size $R \sim 1/m_i$ of a massive non-relativistic body is much bigger than its Compton wavelength $\lambda \sim 1/M$.  Because of this, the mass distribution can be defined through the scalar gravitational formfactor (\ref{trace-form}) instead of the formfactor of $T_{00}$:
\be\label{def-m}
\langle R_{\rm m}^2 \rangle = \frac{6}{M}\ \frac{d G}{d t}\Big|_{t=0} ,
\ee
where we took into account the normalization (\ref{con-m}).
This conclusion agrees with the arguments given above on the basis of weak gravitational field limit of the Einstein equation.
Later we will verify that the compositeness scale of the scalar gravitational formfactor $m_s^2 \equiv G(0)/(dG/dt |_{t=0})$ extracted from the experimental data is indeed much smaller than $4M^2$. 
\section{Scale anomaly of QCD and the mass distribution}

Now that we have stablished that the mass radius of the proton can be extracted from the formfactor of the trace of the EMT, let us discuss how this formfactor can be measured. The key to this is the scale anomaly of QCD. In this theory, quantum effects lead to non-vanishing trace of the EMT even for massless quarks \cite{Collins:1976yq,Nielsen:1977sy}:
\be\label{trace-an}
T \equiv T_\mu^\mu = \frac{\beta(g)}{2g}\ G^{\mu\nu a} G_{\mu\nu}^a + \sum_{l=u,d,s} m_l (1+\gamma_{m_l}) {\bar q}_l q_l + \sum_{h=c,b,t} m_h (1+\gamma_{m_h}) {\bar Q}_h Q_h ,
\ee
where $G^{\mu\nu a}$ is the gluon field strength tensor with color index $a$,  the sum in the second and third terms runs over the light and heavy quark flavors $q_l$ and $Q_h$ with masses $m_l$ and $m_h$ respectively, and $\gamma_m$ are the anomalous mass dimensions. The beta-function of QCD \cite{Gross:1973id,Politzer:1973fx} $\beta(g) = \partial g/\partial (log \mu)$ governs the renormalization group running of the QCD coupling $g$ with scale $\mu$:
\be\label{beta}
\beta(g) = - b \frac{g^3}{16 \pi^2} + ..., \ \ b = 11- \frac{2 n_l}{3} - \frac{2 n_h}{3}
\ee
where the first term in $b$ is due to gluon loops, and the second and third terms are the contributions from light and heavy quark loops.

To determine the mass radius of the proton, we will be interested in the matrix element of the operator (\ref{trace-an}) at small momentum transfer $|t| < 4 m_h^2$; the lightest heavy quark is the charm with mass $m_c \simeq 1.25$ GeV, so this inequality implies $|t| \leq 6.25\ {\rm GeV}^2$. In this kinematical region, heavy quarks appear only in virtual $Q{\bar Q}$ pairs; as a result, the heavy quark part of (\ref{trace-an}) cancels the heavy quark contribution to the gluon term \cite{Shifman:1978zn}:
\be
\sum_{h=c,b,t} m_h (1+\gamma_{m_h}) {\bar Q}_h Q_h \simeq - \frac{2 n_h}{3}\ G^{\mu\nu a} G_{\mu\nu}^a.
\ee
As a result, the trace of the energy-momentum tensor that will determine the mass radius of the proton contains only the contributions from light quarks and gluons:
\be\label{trace-light}
T \equiv T_\mu^\mu = \frac{{\tilde \beta}(g)}{2g}\ G^{\mu\nu a} G_{\mu\nu}^a + \sum_{l=u,d,s} m_l (1+\gamma_{m_l}) {\bar q}_l q_l  ,
\ee
where ${\tilde \beta}$ is the beta-function with $b = 11 - 2 n_l/3 = 9$ for three flavors of light quarks, $u, d$ and $s$. 

It is well known that the chiral limit of massless quarks provides an accurate approximation to the physical world; in this limit, the trace of the EMT (\ref{trace-light}) contains only the gluon term. Therefore, since the forward matrix element of (\ref{trace-light}) according to (\ref{norm-m}) yields the mass of the nucleon, we have to conclude that the mass of the proton in the chiral limit is entirely due to gluons. The contribution of the second term (``$\sigma$-term") in (\ref{trace-light}) for physical values of light quark masses can be extracted from the experimental data on pion and kaon scattering amplitudes (for recent work, see \cite{RuizdeElvira:2017stg}) or computed in lattice QCD \cite{Yang:2015uis}; it contributes about $80\ {\rm MeV}$, or about $8 \%$, to the total proton mass -- so the chiral limit is indeed reasonably accurate.

\section{Quarkonium photoproduction near the threshold}

In the chiral limit, the information about the mass radius of the proton is contained in the matrix element of the scalar gluon operator in (\ref{trace-light}) at non-zero momentum transfer. The zero-momentum transfer, forward matrix element of this operator yields the proton's mass, and this can be used for evaluating the scattering length in quarkonium-nucleon interaction \cite{Luke:1992tm,Kaidalov:1992hd,Kharzeev:1995ij}. 

At finite momentum transfer, the matrix element of the scalar gluon operator in (\ref{trace-light}) can be measured in photoproduction of vector heavy quarkonium states, $J/\psi$ and $\Upsilon$ close to the threshold \cite{Kharzeev:1995ij,Kharzeev:1998bz}. This proposal is based on the following arguments:
\begin{enumerate}
{\item Because $J/\psi$ and $\Upsilon$ are made of a heavy quark and an antiquark, and the proton at small momentum transfer contains only light quarks, the corresponding amplitude is dominated by the exchange of gluons.}
{\item Close to the threshold, the characteristic size of the heavy quark-antiquark pair is $\sim 1/(2 m_h)$; for charm quarks, this is about $0.08\ {\rm fm}$. Because this size is much smaller than the radius of the proton, the coupling of gluons to the heavy quark is perturbative, is characterized by a small coupling constant, and can be described by a local color-neutral gluon operator of the lowest possible dimension \cite{Voloshin:1978hc,Appelquist:1978rt,Peskin:1979va}.}
{\item Because of the vector quantum numbers $J^{PC}=1^{--}$ of $J/\psi$ and $\Upsilon$, the threshold photoproduction is due to the $t-$channel exchange of gluons in scalar $0^{++}$ and tensor $2^{++}$ states; the scalar exchange is described by the operator that is proportional to the first term in (\ref{trace-light}). Because of the scale anomaly, its matrix element does not depend on the QCD coupling constant $g^2$, whereas the matrix element of the tensor operator appears proportional to $g^2$, and is sub-leading at weak coupling \cite{Luke:1992tm,Kaidalov:1992hd,Kharzeev:1995ij,Kharzeev:1998bz,Fujii:1999xn}.}
\end{enumerate}

Let us now examine these arguments in more detail, and use them for the extraction of the mass radius of the proton. Consider the interaction of the heavy quark pair with the proton near the threshold, where the velocity of the quarkonium in the center-of-mass is small: $v_\psi \ll c$. The coupling of a small color-neutral heavy quark-antiquark state to gluons can be described by the operator $g^2 {\bf E}^{a2}$, where ${\bf E}^{a}$ is the chromo-electric field -- this is the quadratic QCD Stark effect (the first-order effect is forbidden by color neutrality). The chromo-magnetic contribution is proportional to $(v_\psi/c)^2$ and is suppressed; the operators that contain covariant derivatives are suppressed by the powers of  $(v_\psi/c)^2$ as well. The $g^2 {\bf E}^{a2}$ operator can be identically represented as a sum of the scalar $0^{++}$ and tensor $2^{++}$ gluon operators \cite{Novikov:1980fa}:
\be\label{chromo-el}
g^2 {\bf E}^{a2} = \frac{g^2}{2} ({\bf E}^{a2} - {\bf B}^{a2}) + \frac{g^2}{2} ({\bf E}^{a2} + {\bf B}^{a2}) = \frac{8 \pi^2}{b}\ T + g^2 T_{00}^{(g)}, 
\ee
where $T$ is the trace of the EMT (\ref{trace-light}) in the chiral limit:
\be
T = \frac{{\tilde \beta}(g)}{2g}\ G^{\mu\nu a} G_{\mu\nu}^a = - \frac{b g^2}{32 \pi^2} G^{\mu\nu a} G_{\mu\nu}^a ,
\ee
and $T_{00}^{(g)}$ is the temporal component of the gluon part of the EMT of QCD.
\vskip0.3cm

\begin{figure}[t]
\centering
\hspace{-1.8cm}\begin{minipage}{.49\textwidth}
\includegraphics[scale=0.25]{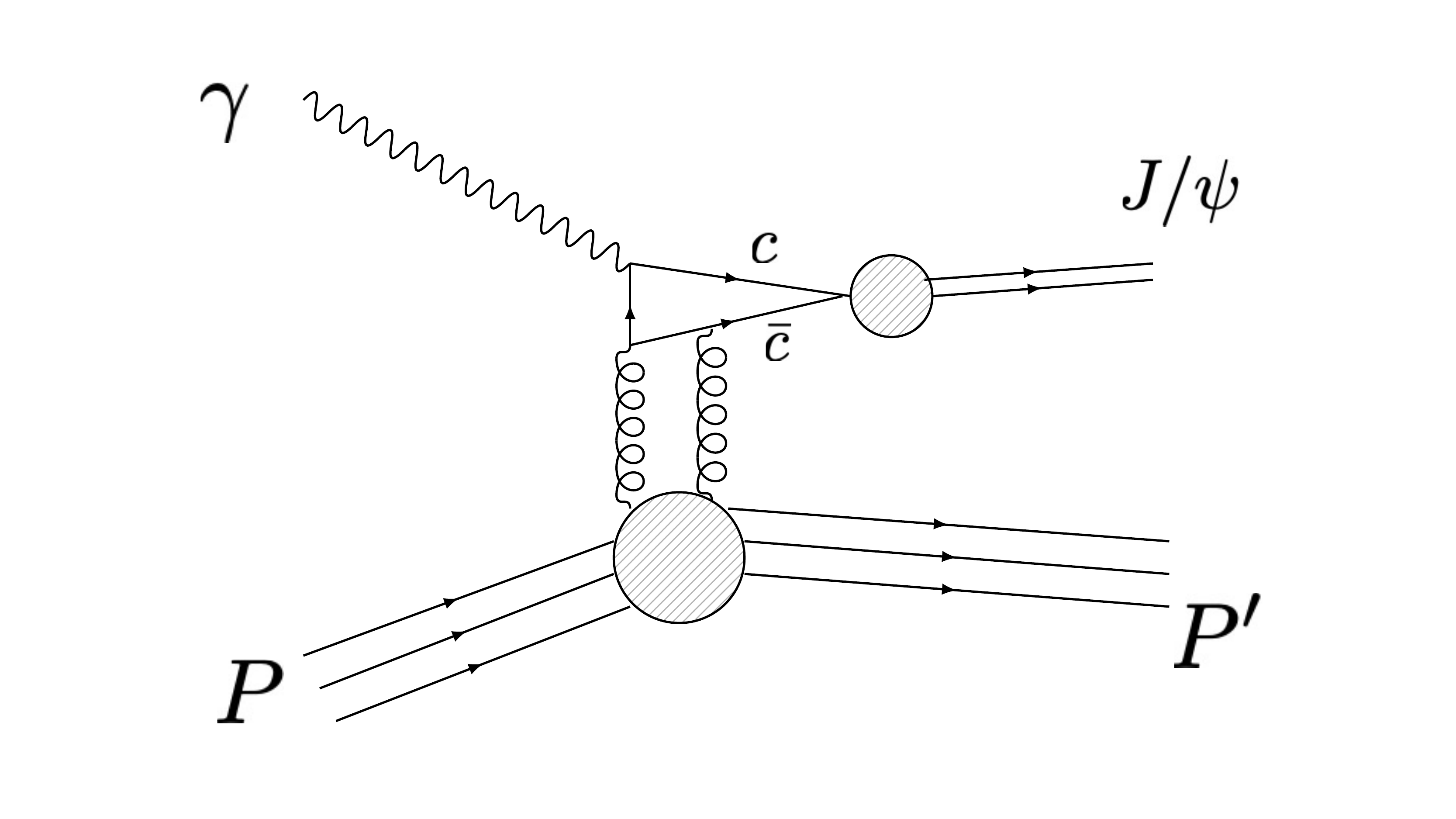}
\label{feyn_diag}
\end{minipage}\
\begin{minipage}{.5\textwidth}
\includegraphics[scale=0.33]{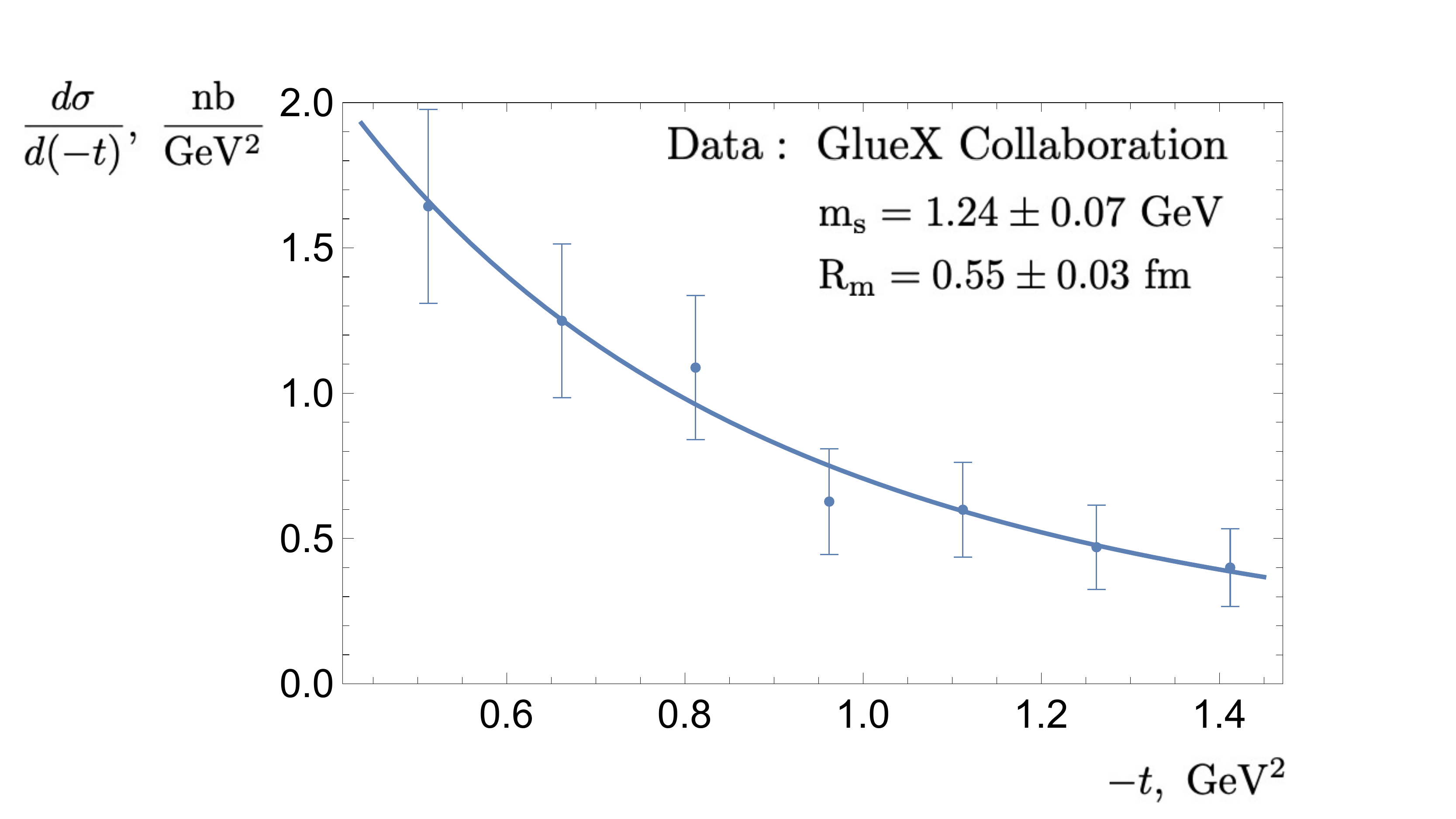}
\label{fit}
\end{minipage}
\caption{Left: the Feynman diagram of $J/\psi$ photoproduction off a proton. Right: the differential cross section of $J/\psi$ photoproduction at the center-of-mass energy $E_{cm} = 4.58$ GeV (lab. energy of the photon $E_\gamma =  10.72$ GeV); the data is from the GlueX Collaboration \cite{ali2019first}; the theory curve corresponds to the dipole form of the scalar gravitational formfactor with the parameter $m_s = 1.24 \pm 0.07\ {\rm GeV}$, corresponding to the mass radius of the proton $R_{\rm m} = 0.55 \pm 0.03$ fm.}
\label{fig:1}
\end{figure}

The amplitude of $J/\psi$ photoproduction close to the threshold factorizes into a short-distance part describing the electric polarizability of the $c{\bar c}$ pair, and the matrix element of the operator (\ref{chromo-el}) over a proton, see Fig. \ref{fig:1} (left):
\be\label{ampl}
{\cal M}_{\gamma P \to \psi P}(t) = - Q e\ c_2\  2M\ \langle P'| g^2 {\bf E}^{a2} | P \rangle ,
\ee
where $Q e = 2e/3$ describes the coupling of the photon to the electric charge of the charm quark, $c_2$ is the short-distance coefficient describing the coupling of the chromoelectric fields to the heavy quark pair, and its transition to the $J/\psi$,  
$t = (P'-P)^2$ is the momentum transfer, and the factor $2M$ is needed to reconcile the relativistic normalization of states with our normalization of the EMT formfactors (\ref{mass}, \ref{norm-m}). 

The expression (\ref{ampl}) holds only near the threshold, where the scalar gluon operator dominates over the operators that contain derivatives; they are suppressed by powers of $J/\psi$ velocity squared, $(v_\psi/c)^2$. Note that in this kinematical region the scalar gluon formfactor that enters (\ref{ampl}) cannot be interpreted in terms of the gluon structure functions -- indeed, the gluon structure functions are defined through the matrix elements of {\it traceless} gluon operators in the Operator Product Expansion, and we are interested in the trace part. 

Substituting the relation (\ref{chromo-el}) in (\ref{ampl}), we observe that the matrix element of the first term does not contain the coupling $g^2$ as a consequence of scale anomaly, whereas the second term in (\ref{chromo-el}) is suppressed by $g^2$ which is small at the scale of $Q^2= 4 m_c^2$; in addition, the first term is enhanced by a numerical factor. Therefore we can re-write the amplitude (\ref{ampl}) in terms of the scalar gravitational formfactor:
\be\label{ampl1}
{\cal M}_{\gamma P \to \psi P}(t) = - Q e\ c_2\  \frac{16 \pi^2 M}{b}\ \langle P'| T | P \rangle .
\ee
The differential cross section and the integrated cross section of the $J/\psi$ photoproduction can now be computed using the standard formulae, see e.g. \cite{particle2020review}:
\be\label{cross}
\frac{d \sigma_{\gamma P \to \psi P}}{dt} = \frac{1}{64 \pi s}\ \frac{1}{\left|{\bf p}_{\gamma cm}\right|^2}\ \left| {\cal M}_{\gamma P \to \psi P}(t) \right|^2 ,
\ee 
and
\be\label{cross-int}
\sigma_{\gamma P \to \psi P}(s) = \int_{t_{min}}^{t_{max}} dt\ \frac{d \sigma_{\gamma P \to \psi P}}{dt} ,
\ee
where ${\bf p}_{\gamma cm}$ is the photon momentum in the c.m.s. of the process, and $s=(p_\gamma+P_p)^2$ is the square of the c.m.s. energy.
We expect that the short-distance coefficient $c_2$ is on the order of $\pi r_{c{\bar c}}^2$, where the size of the $c{\bar c}$ pair $r_{c{\bar c}} \simeq 1/2m_c \simeq 0.08$ fm. We will fit this parameter to the GlueX Collaboration data \cite{ali2019first}, and then check that it is in the expected range.
\vskip0.3cm

Let us briefly discuss the kinematics of the $\gamma + p \to J/\psi + p$ process. Because of the large mass of $J/\psi$, close to the threshold the process is characterized by a sizable minimal momentum transfer $t_{min}$; right at the threshold, its value is $t_{min} = - M_\psi^2 M/(M_\psi + M) \simeq - 2.23\ {\rm GeV}^2 \simeq - (1.5\ {\rm GeV})^2$, where $M_\psi \simeq 3.097$ GeV is the mass of $J/\psi$ and $M \simeq 0.938$ GeV is the mass of the proton. The large magnitude of $t_{min}$ close to the threshold makes the use of the vector meson dominance model questionable. On the other hand, $t$ is still much smaller than $4m_c^2 \simeq 6.25\ {\rm GeV}^2$ which justifies the approach based on Eqs (\ref{ampl1}) and (\ref{cross}). In this kinematical domain, the factor $c_2$ in (\ref{ampl1}) can indeed be treated as a constant; when the magnitude of $t$ becomes comparable to $4m_c^2 \simeq 6.25\ {\rm GeV}^2$, this factor can be expected to acquire a significant $t$ dependence. Because $t_{min}$ rapidly varies with the c.m.s. energy close to the threshold, the energy dependence of the integrated cross section (\ref{cross-int}) is sensitive the scalar gravitational formfactor. However the quantity that is most sensitive to the scalar gravitational formfactor is the differential cross section (\ref{cross}). 

The dominance of the scalar gluon operator over the operators with covariant derivatives in the QCD multipole expansion is justified by the smallness of the heavy quark pair velocity in the c.m.s., $v_\psi$. The operators with derivatives are suppressed by even powers of  $v_\psi$ - therefore, to limit their contributions by less than about $10\%$, we have to limit $v_\psi \leq 0.3$ -- this translates into the limit on the c.m.s. energy $E_{cm} \leq 4.6$ GeV, or the photon lab. frame energy of $E_\gamma \leq 11.3$ GeV.

\section{Extracting the mass radius of the proton}

To make a direct comparison with the charge radius of the proton which has been traditionally extracted by using the dipole form factor \cite{Feynman:1973xc}, we will assume, as a first step, a simple dipole parameterization for the scalar gravitational formfactor as well:
\be\label{dip}
G(t) = \frac{M}{\left(1 - \frac{t}{m_s^2}\right)^2} ,
\ee
where $m_s$ is the only adjustable parameter. With the standard definition (\ref{def-m}), this parameter relates to the r.m.s. mass radius of the proton in the following way:
\be
\langle R_{\rm m}^2 \rangle = \frac{12}{m_s^2} .
\ee
\vskip0.3cm

Using (\ref{ampl1}) with the formfactor $\langle P' | T | P \rangle = G(t)$ given by (\ref{dip}) to evaluate the differential cross section (\ref{cross}), we can now perform the fit of the recent data from the GlueX Collaboration \cite{ali2019first} at Jefferson Lab that is available at $E_\gamma \simeq 10.72$ GeV, 
which is within our desired kinematical range as described above. The resulting fit of the data is shown in Fig. \ref{fig:1} (right); the extracted value 
\be\label{sc-mass}
m_s = 1.24 \pm 0.07\ {\rm GeV}
\ee
 corresponding to 
\be\label{rad-val}
R_{\rm m} \equiv \sqrt{\langle R_{\rm m}^2 \rangle} = 0.55 \pm 0.03\ {\rm fm}
\ee
provides an excellent fit with adjusted ${\bar r}^2 = 0.99$ (${\bar r}^2 = 1.0$ implies a perfect fit). 

The corresponding value of the short-distance coefficient in (\ref{ampl1}) describing the coupling of the gluons to the heavy quark pair is $|c_2|^2 = 0.043 \pm 0.006\ {\rm fm}^4$. Each gluon couples to the chromoelectric dipole moment of the the heavy quark pair, and so $c_2$ has dimension of length squared. As discussed above, we expect that $c_2$ is on the order of $\pi r_{c{\bar c}}^2$, where the size of the $c{\bar c}$ pair $r_{c{\bar c}} \simeq 1/2m_c \simeq 0.08$ fm. The extracted value of $|c_2|^2 = 0.043 \pm 0.006\ {\rm fm}^4$ corresponds to $r_{c{\bar c}} \simeq 0.1$ fm, which is in line with our expectations.

We can compare the value of the mass radius (\ref{rad-val}) to the proton charge radius \cite{particle2020review}
\be
R_{\rm C} \equiv \sqrt{\langle R_{\rm C}^2 \rangle} = 0.8409 \pm 0.0004\ {\rm fm}
\ee
that is known with a much better precision, see \cite{Bernauer:2020ont} for a recent review. 
It appears that the mass radius of the proton is about $50 \%$ smaller than its charge radius! This observation is statistically significant given the statistical error bar of our value (\ref{rad-val}). Of course, there is also a theoretical systematic  uncertainty that we will discuss below; however it does not appear large enough to explain the observed difference between the charge and mass radii of the proton.
\vskip0.3cm
\section{The mass radius puzzle}

\begin{figure}[t]
\centering
\hspace{-1cm}
\includegraphics[scale=0.4]{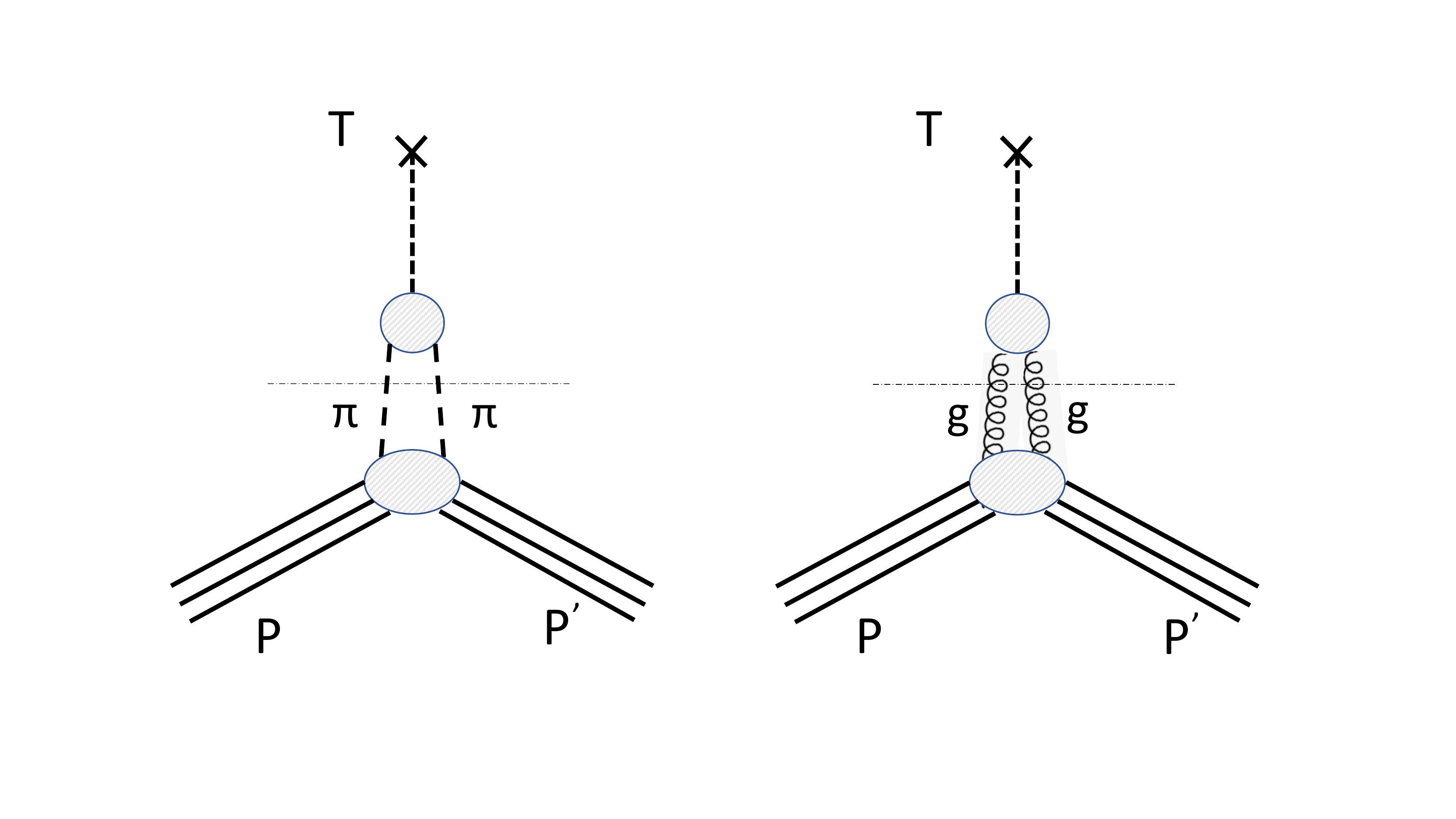}
\caption{Spectral representation of the scalar gravitational formfactor. Left: large-distance, small invariant mass region, where the dominant contribution is from the pion pair. Right: short-distance, large invariant mass region that is dominated by gluon pairs.}
\label{fig:2}
\end{figure}

At first glance, this difference may seem surprising -- but only if one thinks of a proton as of a charged ball of a fixed radius, and not as of a quantum object. First, the charge radius is extracted from the coupling of the photon to quarks, whereas the mass radius results from the coupling to gluons -- and it appears that the gluon radius of the proton is significantly smaller than its quark radius. If we write down the spectral representation for the scalar and charge formfactors (see Fig. \ref{fig:2}), the radii of the mass and charge distributions can be seen to be set by the masses of the lightest physical states excited from the vacuum by the scalar gluon and vector quark currents, correspondingly. For the vector quark current, the lightest physical state is the $\rho$ meson with mass of $m_\rho \simeq 770$ MeV, whereas for the scalar gluon current it is the scalar glueball, with a much larger mass of $m_G \simeq 1600$ MeV, see e.g. \cite{athenodorou2020glueball}. 

While this argument does explain a much smaller mass radius, it is too na\"ive - in the QCD vacuum, the scalar gluon current strongly mixes with the scalar quark one. In fact, there exists a low energy theorem \cite{Novikov:1981xi} that can be used to relate the off-diagonal correlation function of the scalar gluon and quark currents to the quark condensate. The analysis \cite{Fujii:1999xn,Ellis:1999kv} based on this low energy theorem and the experimental hadron spectrum shows a very strong mixing of the scalar glueball state with the $f_0(980)$ meson and a broad $\sigma(500)$ resonance in the $\pi\pi$ spectrum, which is lighter than the $\rho$ meson. Therefore, the hadron spectrum  in the scalar  and vector channels alone cannot explain the difference between the mass and charge radii. 

The reason for the smallness of the mass radius in our opinion is the interplay of scale anomaly and spontaneously broken chiral symmetry. The dominant contribution at large distances in the dispersion representation of the proton formfactor stems from the scalar $\pi\pi$ exchange, see Fig. \ref{fig:2}, left. 
Because the trace of the EMT is invariant under the Renormalization Group (RG), its matrix element at small momentum transfer (responsible for the long-range tail of the mass distribution and thus for the mass radius) can be calculated \cite{Voloshin:1980zf} using the effective chiral theory. Matching onto the chiral perturbation theory allows to evaluate the matrix element of the trace of the EMT in a model-independent way; in the chiral limit $T= - (\partial_\mu \pi)^2 + ...$, so we get \cite{Voloshin:1980zf}
\be\label{match}
\langle 0 | T | \pi^+ \pi^- \rangle = q^2,
\ee
where $q^2$ is the invariant mass of the $\pi\pi$ pair. In QCD with $N_f$ light quark flavors, the r.h.s. of (\ref{match}) gets multiplied by $(N_f^2-1)$ after summing over all possible pion combinations, so the corresponding spectral density of the correlation function of $T$ is 
\be\label{spect}
\rho^{\pi\pi}(q^2) = \frac{N_f^2-1}{32 \pi^2}\ q^4 .
\ee
The matrix element (\ref{match}) and the spectral density (\ref{spect}) are strongly suppressed at small invariant masses, corresponding to long-range tail of the proton mass distribution -- as a result, the spectral density of the correlation function of $T$ peaks at masses about $\sim 1$ GeV \cite{Fujii:1999xn}, consistent with  our finding (\ref{sc-mass}).  The underlying reason for the suppression of the matrix element (\ref{match}) at low $q^2$ is the fact that Goldstone bosons decouple from the scalar curvature induced by the scale anomaly -- this is analogous to the absence of light bending in scalar gravity discussed in Section \ref{sec:grav}.

At short distances, where the invariant masses in the spectral representation are large, the relevant matrix element of the trace of the EMT can be computed using the QCD perturbation theory; in this case, the two-gluon state dominates (see Fig. \ref{fig:2}, right), with the spectral density \cite{Fujii:1999xn}
\be\label{match-pt}
\rho^{pert}(q^2) = \left(\frac{b g^2}{32 \pi^2}\right)^2\ \frac{N_c^2-1}{4 \pi^2}\ q^4 ,
\ee
where $b=(11 N_c - 2 N_f)/3$ is the coefficient of the QCD beta-function. We thus see that the mass distribution at short distances is governed by the Renormalization Group flow, and at large distances -- by the interplay of scale anomaly and spontaneously broken chiral symmetry. Because pions are Goldstone bosons, their couplings involve derivatives of the pion field -- otherwise they would not be invariant w.r.t. the chiral rotations. Because of this, at small momenta pions decouple from the trace of the EMT, and this confines the mass distribution of the proton to shorter distances, where it is dominated by the RG flow of QCD.

It would be interesting to decompose the mass distribution into the quark and gluon ones. The mass decomposition for the proton \cite{Ji:1994av,Kharzeev:1995ij,Yang:2018nqn,Polyakov:2018zvc,Lorce:2017xzd,Metz:2020vxd,Alexandrou:2020sml} is a subject of a lively debate at present, and is subject to the renormalization scale and scheme dependence, as well as to the frame dependence in the case of $T_{00}$. When quark and gluon contributions are separated, an additional term appears in the decomposition (\ref{gen}) that corresponds to the anomalous gravimagnetic moment and gravielectric dipole moment. These terms however should cancel each other in the total TEM of the nucleon to obey the Einstein equivalence principle, and this is why we did not consider them.

The formfactor of the gluon term in the TEM decomposition has been recently evaluated in lattice QCD \cite{shanahan2019gluon}. The authors fit this formfactor by the dipole form, and extract the effective mass of $m_s =   1.13 \pm 0.06$ GeV -- remarkably, this is consistent with the value $m_s = 1.24 \pm 0.07$ that we have extracted above from the GlueX data \cite{ali2019first}.
The scalar and tensor gravitational formfactors that enter the photoproduction amplitude have also been evaluated in the approaches based on holography \cite{hatta2018holographic,hatta2019near,mamo2020diffractive}. In this case, the scale of the formfactor is encoded in the dilaton potential in the bulk that is constructed to reproduce the hadron spectrum and Regge trajectories. 

A number of papers address the proton mass decomposition basing on the vector meson dominance \cite{Kharzeev:1995ij,Kharzeev:1998bz,gryniuk2016accessing,strakovsky2020j,gryniuk2020Upsilon,wang2020origin}. Near the threshold, the scattering amplitude of quarkonium possesses a large real part directly related to the scale anomaly that strongly affects the cross section, which thus becomes sensitive to the mass decomposition of the nucleon \cite{Kharzeev:1995ij,Kharzeev:1998bz}. However a big problem of this approach stems from a large value of $t_{min}$ near the threshold that necessitates taking account of excited $c{\bar c}$ states, with unknown and interfering scattering amplitudes. 

Perturbative approaches to the threshold photoproduction of quarkonium introduce the two-gluon formfactors and relate them to gluon distributions \cite{Frankfurt:2002ka,zeng2020near}. While these approaches are close to ours in terms of phenomenology, we stress that the scalar gravitational formfactor that we have considered cannot be interpreted in terms of gluon structure functions, that are defined through the matrix elements of traceless gluon operators; instead, they originate from the trace operator that dominates near the threshold. In the language of Operator Product Expansion (OPE), the trace terms correspond to the target mass corrections to the parton model, see e.g. \cite{Kharzeev:1996tw}. Non-perturbative instanton contributions to scalar formfactors have been recently evaluated in \cite{Shuryak:2020ktq}.

The contribution from $\Lambda {\bar D}^*$ $t$-channel exchanges has been studied in \cite{du2020deciphering}; in our approach based on the OPE,  this contribution corresponds to higher dimensional operators that should be suppressed by the heavy quark mass. \vskip0.3cm
\section{Theoretical uncertainties and an outlook}
The error bar in the mass radius (\ref{rad-val}) originates only from the precision of the current data \cite{ali2019first}. But what is the ``theoretical systematic error" involved in its extraction? We can categorize the sources of the uncertainty in (\ref{rad-val}) as follows:
\begin{itemize}
{\item The contribution of gluon operators with derivatives: as mentioned above, the contamination from these operators is suppressed by powers of the $J/\psi$ velocity in the c.m.s., $v_\psi$. The GlueX data \cite{ali2019first} on the differential cross section that we used are at the energy of $E_\gamma \leq 11.3$ GeV, corresponding to $v_\psi \simeq 0.3$. This means that the potential contribution of operators with derivatives is about $10 \%$. This contribution can be further reduced by measuring the differential cross section at a lower energy -- for example, at $E_\gamma \leq 9.4$ GeV it should be less than $5 \%$.
}
{\item Operators of higher dimension, e.g. the quark-gluon operators: the contribution from these operators is suppressed by the powers of $\Lambda^2/(4 m_c^2)$, where $\Lambda$ is a non-perturbative parameter. Assuming $\Lambda \sim 1$ GeV, we expect these contributions to be on the order of $\sim 10 - 15\ \%$. Note that in threshold photoproduction, the convergence of the multipole expansion is much better than for the interaction of on-shell quarkonia where it is governed not by the heavy quark mass but by the quarkonium binding energy \cite{Peskin:1979va,Kharzeev:1995ij}.
The contributions of higher-dimensional operators can be further suppressed by extending the studies of threshold photoproduction to the $\Upsilon$ states, that should become possible with the advent of the Electron Ion Collider. It could also be possible to study this process in ultra-peripheral collisions at RHIC and LHC.
}
{\item The extrapolation in momentum transfer $t$ and the $t$-dependence of the short distance coefficient $c_2$: we have assumed that the coefficient $c_2$ is $t$-independent. This is justified by the fact that the momentum transfer in the threshold photoproduction is not large enough to resolve the internal structure of the produced heavy quark pair with the size $\sim 1/(2 m_c)$, $t \ll 4 m_c^2$. The range of extrapolation to $t=0$ from $t_{min}$ is also  much smaller than $4 m_c^2$, which justifies the assumption of a constant $c_2$. }
\end{itemize}

A careful evaluation and reduction of the uncertainty in (\ref{rad-val}) will require a lot of dedicated theoretical and experimental effort. However the mass distribution is definitely a fundamental property of the proton. Therefore the measurements of this distribution, combined with measurements of other ``mechanical" properties of the proton, such as the pressure distribution \cite{Burkert:2018bqq}, will definitely advance our understanding of the quantum origin of mass.
\vskip1.8cm

{\bf Acknowledgements}

This work was first presented at the ``Origin of the Proton Mass" workshop organized at ANL by I. Clo\"et, X. Ji, Z. Meziani, and J. Qiu; I am grateful to the organizers and participants of this workshop for stimulating discussions. I also thank J. Bernauer and {\mbox{ Z. Meziani}} for useful comments on the manuscript.
This research was supported by the U.S. Department of
Energy under awards DE-FG88ER40388 and DE-SC0012704.

\bibliography{mass}

\end{document}